# A QCA Layout Design Methodology. Part I

Shadi Sheikhfaal

Department of Electrical and Computer Engineering, University of Central Florida, Orlando, FL 32816

*Abstract*—**Quantum-dot Cellular Automata (QCA) as a nanoscale transistor-less device technology offers distinguishing advantages over the limitations of CMOS circuits. While more than 2 decades of design progress have been achieved with QCA, a comprehensive composition approach for the layout design in this technology is needed. In this study, the *Priority-Phased Decomposition-Driven (PPDD)* QCA logic design methodology is proposed. In this methodology, after partitioning combinational and sequential circuits into *primary-level priority blocks* including 2:1 MUX and XOR gates, and *secondary-level priority blocks* comprising multi-input majority gates, there are three streamlined approaches which are developed to compose the desired QCA circuit using blocks from each priority level.**

*Index Terms*— **Quantum-dot cellular automata (QCA), Decomposition method, QCA layout, Majority Gate, Multiplexer Design.**

## I. INTRODUCTION

Over the last decade, the demands of high speed operation with low power consumption along with the scaling constraints of CMOS structures have compelled the scientists to investigate alternative emerging technologies [1]. Therefore, many technologies have been investigated such as Carbon Nano Tube field effect transistors (CNTFETs), Single electron transistors (SETs), Quantum-dot cellular automata (QCA) and others. QCA is a promising technology supporting a transistor-less paradigm. In this nanoscale technology, circuits operate with high speed switching characteristics, extremely high densities, and at low power consumption [2], [3], all of which are sought for next generation circuits specifically memory devices. QCA faces challenges related to fault occurrence during fabrication, ambient conditions of operation, and circuitry layout, which we address the latter concern by developing a new decomposition and clocking approach.

Efficient bit storage cell design occupies a pivotal role in the advancement of QCA technology. It is determined by the logical building blocks utilized, as well as the interconnect requirements that are incurred. The building blocks are determined by the fundamental logic gate structure. Each QCA cell contains two electrons and four quantum dots and due to Coulomb interaction between these identical charges, they occupy dots diagonally. As a result, the two stable polarization states for a QCA cell are achieved, as shown in Fig. 1(a). The instantaneous polarization of a cell is denoted as either -1 or +1, which are encoded to represent a binary "1" value and "0" value, respectively. Meanwhile, the interconnect requirements are determined by the topology and interactions between cells. A chain of QCA cells placed side-by-side is capable of propagating the first cell's polarization via successive cell interactions to realize the role of a conductor in CMOS devices [4]. Fig. 1(b) demonstrates a QCA wire propagating binary "0" value.

In QCA technology, storage cells do not require an external power source to maintain their current stable polarization. Actually, the clock controls the flow of charge in the circuit. The QCA clock consists of four clock phases, i.e. *Switch*, *Hold*, *Release*, and *Relax*, which span a 90 degree out-of-phase progression [5], [7]. Recently, an extensive QCA power consumption model has been presented in [6]. In this model, the total power of QCA circuits is divided into two main elements, leakage power and switching power. Power loss during clock fluctuations, i.e. from low to high or high to low, is designated as leakage power and power loss during switching period is designated as the switching power [5].

Recently, studies have also been undertaken to design diverse QCA structures more efficiently. For example, there are several optimized designs for QCA adders [1], [8], [9],[48] [31], [40],[41],[44]-[46], multiplexers [9]-[14], logical circuits using exclusive-OR gates [5], [14], reversible gates [42],[43],[45] and various alternative QCA wiring approaches [8],[18]. Moreover, much attention has been paid to the memory cell design as a vital elements in the QCA devices [15], [37], [47]. QCA loop-based approaches constitute a pioneering method to designing well-optimized with the ambitious objective to augment traditional CMOS-based Static Random Access Memory (SRAM) [16], [17], [19], [20].

While techniques have been established to design each of these circuits individually, the needs for a general design methodology capable of realizing arbitrary QCA architectures are identified in the research literature. In [35], a new design method has been presented that exploits in new ways the properties of auxiliary propagate and produces signals in order to diminish the majority gate numbers required to implement QCA adders. The authors of [36] indicate a novel methodology for error analysis in QCA circuits using deterministic and random insertion of possible defects. In [37], a modular technique has been proposed that relies on QCA tiles.

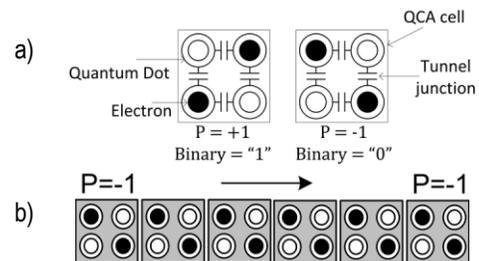

Fig. 1. (a) Basic QCA cell with two binary state (b) QCA standard wire.

In this study, a comprehensive composition methodology for designing QCA circuits is proposed. In this methodology, after dividing combinational and sequential circuits into two levels of priority blocks (first priority blocks including 2:1 MUX and XOR gates and second priority blocks comprising multi-input majority gates), three efficient approaches are employed in order to implement the target QCA circuit. These three approaches are entitled conventional approach, innovative approach, and cell level method.

## II. BACKGROUND

### A. QCA Fundamental Concepts

Inverters and majority gates are basic elements in QCA-based design. The exact output polarization in these gates is computed using cell interactions through kink energy. This energy is associated with the energy cost of two QCA cells ($i$ and $j$) having opposite polarizations and is expressed as [5,38]:

$$E_{i,j} = \frac{1}{4\Pi\varepsilon_0\varepsilon_r} \sum_{n=1}^{4}\sum_{m=1}^{4} \frac{q_{i,n}q_{j,m}}{\left|r_{i,n}-r_{j,m}\right|} \quad (1)$$

As is shown in Fig. 2 (a), an inverter gate can be constructed using two cells diagonally placed next together. The next primary gate used in QCA logic designs is the majority gate [2]. As is clear in Fig. 2 (b), the majority of the inputs' logic is transferred to the output cell. Previously, a highly-used design for three-input majority has been presented. Assuming A, B and C as input cells, the output function is AB+AC+BC [3]. Two-input AND gate and two-input OR gate are made by applying fixed polarization to one of the inputs. For example, if input C is fixed to -1, the three-input majority gate acts like a two-input AND gate. Correspondingly, by fixing input C to +1, the three-input majority gate becomes a two-input OR gate.

Attempts for expanding the majority gate concept to exploit its beneficial low-complexity structure led to theorizing an innovative design for five-input majority gate [5]. This structure as is plotted in Fig. 2(c), can facilitate making three-input AND/OR gates with the least area occupation. The five-input majority gate equation can be written as:

$$M(A,B,C,D,E) = ABC + ABD + ABE + ACD + ACE + ADE + BCD + BCE + BDE + CDE \quad (2)$$

One of the challenges in QCA technology is the need to route orthogonal signal channels, which may incur crossing of QCA wires. To address this need, several solutions have been advocated. The first method presented in [1] utilizes two types of QCA cells, *standard* and *rotated*. As is illustrated in Fig. 3(a), in the coplanar wire-crossing, one of the wires is made by standard cells and other one is made by rotated cells. In this case, the two wires operate independently and do not have any effect on each other. Another approach presented in [8] is named logical wire-crossing. In this method, by considering a 180 degree out-of-phase assignment, then the cells placed in the hold phase can allow polarization to cross over the cells placed in the relax phase. Similarly, crossover of cells in the switch phase and the release phase can be readily performed. Implementation of this approach is shown in Fig. 3(b).

It is noteworthy that QCA synchronization concepts enforce designers to consider some design rules in their implementations for obtaining precise functioning and reducing circuit's susceptibility [3], [32], such as minimum and maximum cell number in a clock zone or the need to synchronize flows arriving to a gate.

Based on [9,32,39], each QCA clock zone should consist of at least two cells to maintain its influence during the next clock zone. Moreover, majority gate's synchronization is accomplished once all the inputs are assigned to integral structure of a multi-input majority gate (as an instance 5-cell scheme of 3-input majority gate) in a specific clock zone right before majority gate's clock zone. Accordingly, it is mentioned that the output wire connecting to next gate should be positioned at a clock zone after middle cells. This robust structure guarantees high polarized signal transmission.

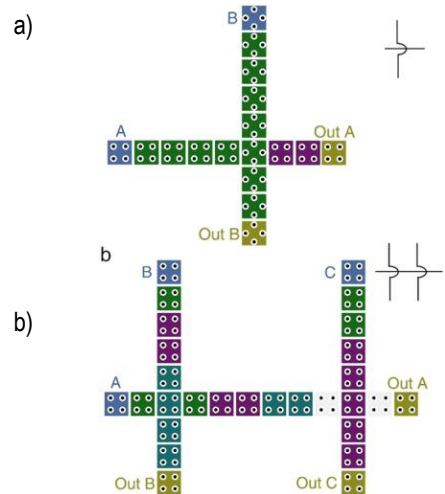

Fig. 3. (a) Coplanar wire-crossing, (b) Logical wire-crossing (Clocking based).

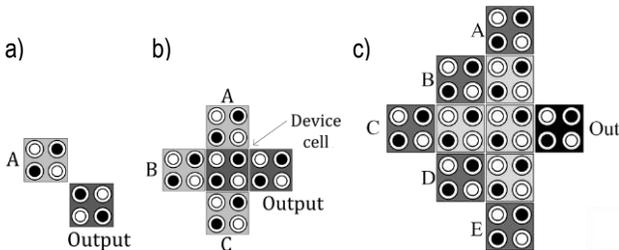

Fig. 2. (a) QCA inverter (b) Three-input majority gate (c) Five-input majority gate.

### III. QCA DESIGN METHODOLOGY FOR DESIGNING EFFICIENT LAYOUTS (SRAM CELL CASE STUDY)

In this section, a structured and consistent methodology for designing quantum-dot cellular automata circuits is proposed which is named Priority-Phased Decomposition-Driven (PPDD). Without loss of generality, a QCA circuit composition, combinational and sequential, can be explored in two levels or phases of abstraction categorized as:

1) *Primary-Level priority blocks* or large-scale ones such as 2- and 3-input Exclusive-OR gates or 2:1 multiplexer circuit. These blocks are regularly constructed utilizing number of QCA multi-input majority gates interconnecting together. The main insight behind selecting these logical functions as primary-level priority blocks are their capability to readily implementing all the complex logics either directly (Multiplexer design [34]) or indirectly using connecting AND-OR gates (Exclusive-or Sum of Product or ESOP design [8,33]).

2) *Secondary-Level priority blocks* or fundamental gates such as multi-input majority gates. These blocks play an indispensable role in connecting primary-Level priority blocks.

Fig. 4 depicts the proposed design methodology in flowchart format. In the first step, after selecting the target circuit, logic functions are partitioned into two priority levels of abstraction based on aforementioned concepts. As a case study in this research, QCA Static Random Access Memory (SRAM) cell is examined. Fig. 5 shows a new enhanced and amended schematic diagram of QCA SRAM cell comprising Primary-Level and Secondary-Level elements for implementing loop-based memories. This design incorporates one three-input and one five-input majority gate as Secondary-Level priority QCA building blocks (indicated by circles) in addition to a 2:1 multiplexer block as a Primary-Level priority QCA building block (indicated by a rectangle).

Before describing the detailed steps comprising the methodology, the functionality of the circuit in Fig. 5 is overviewed. A write operation can be accomplished by setting R/W signal to '1' while the *Enable* signal is also set to '1'. During the write operation, the input data, the second input of 2:1 multiplexer, will be transmitted to the last level five-input majority gate where the *Enable* signal as well as two binary '0's and inverted *W/R* signal are other voters. Accordingly, the data is written in the memory loop indicated by the channel labeled *Mloop*, after propagating through the multiplexer unit. Moreover, the read operation can be achieved by setting *W/R* signal to '0' where *Enable* is set to '1', consequently the stored data will be retrieved through a feedback loop and can be achieved from the output. The complete operation of the proposed logical circuit is indicated in Table I. For further details, previous QCA memory cells which are based on a QCA clocking principles are considered in [17-20].In the second step, the Primary-Level priority blocks should be taken into consideration. There are two entirely different PPDD design methods that can be employed by designers as referred to *Gate level method* and *Cell level method* in the presented flowchart in Fig. 4.

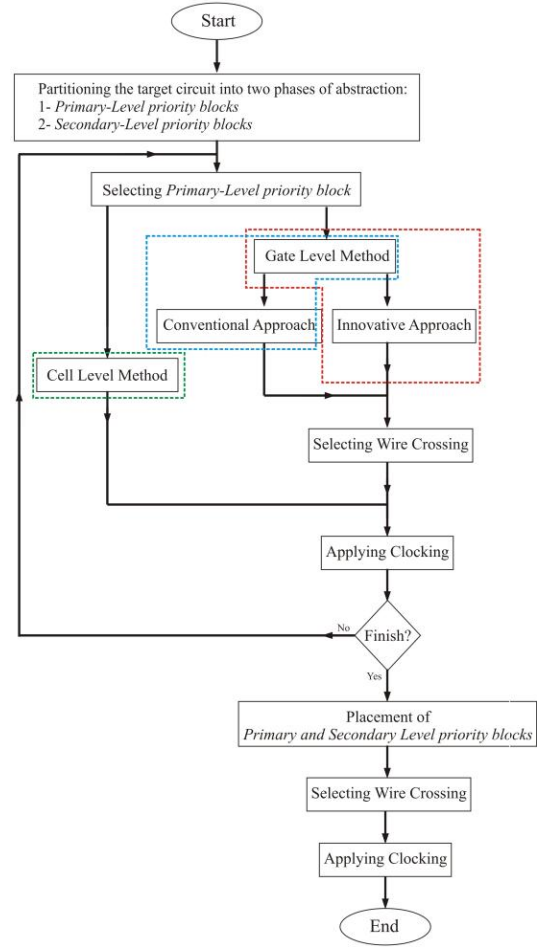

Fig. 4. Priority-Phased Decomposition-Driven methodology steps.

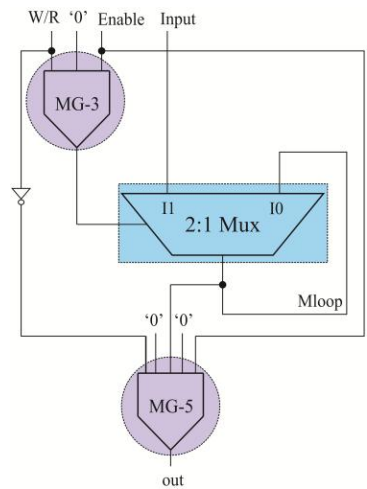

Fig. 5. Logical diagram of the proposed memory cell.

TABLE I
COMPLETE OPERATION OF THE PROPOSED QCA MEMORY CELL

| Operation | Enable | W/R | Input | Mloop | out |
|-----------|--------|-----|-------|-------|-----|
| Write | 1 | 1 | 1 | 1 | 0 |
|  | 1 | 1 | 0 | 0 | 0 |
| Read | 1 | 0 | x | 0 | 0 |
|  | 1 | 0 | x | 1 | 1 |
| Idle | 0 | x | x | Doesn't change | 0 |

### A. Gate level Method of PPDD

In this design method, the designer should propose a new compatible gate-level priority building blocks using multi-input majority gates and then attempt to precisely implement it. QCA compatibility can be defined as the majority gate-based QCA designs which comparatively disregards ubiquitous fully AND-OR based structures. After gate level design, two dissimilar approaches can be exercised, referred to as *conventional* and *Innovative* approaches, hereafter. These approaches are illustrated using blue and red dotted blocks in Fig. 4. The conventional approach utilizes orthodox structures of multi-input majority gates, as described in detail within Section II. The most conspicuous designer's challenge in this approach is to achieve the most efficient implementation by correctly alignment of these gates. Contrary to this prevalent layout design method, in Innovation approach of PPDD method, regardless of conventional structures, new illustrations of fundamental gates can be shown to facilitate the circuit implementation.

Considering a 2:1 multiplexer block as a primary-level block, then three alternative QCA gate-level architectures are considered. In the first one, shown in Fig. 6(a), a 2:1 multiplexer is designed using three three-input majority gates in which two AND gates and one OR gate generate the desired output. Although the last two proposed designs shown in Fig. 6(b) and Fig. 6(c) are also composed of multi-input majority gates, they can be considered as QCA-compatible designs due to their new logical equations. The design in Fig. 6(b) is composed of just one AND along with one OR gate connected to a three-input majority gate. Also, the design in Fig. 6(c) is composed of only two multi-input majority gate completely ignoring AND-OR based designs.

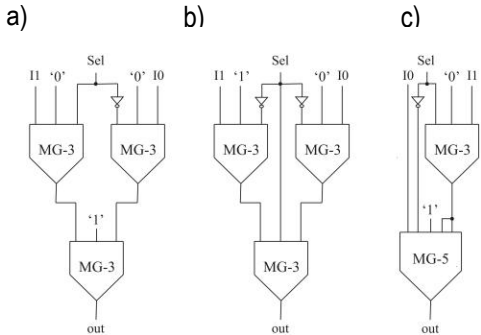

Fig. 6. QCA gate-level designs for *2:1* multiplexer (a) first design using 3 three-input majority gates (b) QCA-compatible design using three-input majority gates (c) alternate QCA-compatible design using three-input and five-input majority gates.

For example, considering the QCA-compatible design shown in Fig. 6(c), this layout can be generated as shown in Fig. 7 [14].

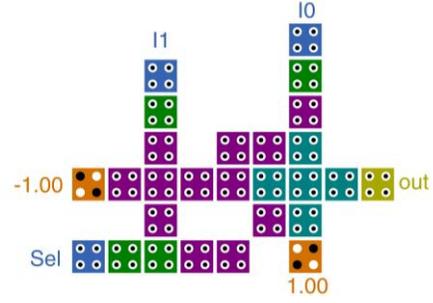

Fig. 7. Proposed QCA layout (M1) of the *2:1* multiplexer employing Gate level method based on the design shown in Fig. 6(c).

According to the flowchart illustrated in Fig. 4, after selecting appropriate Gate-level design and requisite QCA structure, it is vital to choose an adequate wire crossing to interconnect structures. As is clear in Fig. 7, targeting to reduce circuit complexity, we have selected suitable interconnections between majority gates.

### B. Cell level Method of PPDD

In the second design method of PPDD, Primary-Level priority building block is implemented using a novel cell arrangement and without relying on conventional scheme of QCA circuit. Cell level method uses explicit interactions of QCA cells to accomplish the circuit's functionality. However, this implementation method will probably be time consuming, the generated layout has superior features in comparison to Gate level method implementations. Actually, in this method there is no obligation to use of basic QCA structures to acquire correct functioning.

An extensive comparison between previously published QCA 2:1 multiplexers and our proposed layouts based on different methods will be performed in Part II of this study. The latency, cell count, cross-over type, and area occupations of each circuit will be evaluated as important assessment metrics. In the next step, after implementing the first instance of primary-level priority building blocks and applying well-organized clocking, other first priority blocks should be designed and implemented using Gate-level or Cell-level methods of PPDD. By inserting QCA layout of first priority structures in conjunction with second priority blocks, initial layout of the target circuit will be delineated. Finally, a well-adapted wire crossing approach and suitable clocking should be applied to the generated layout. In our case study presented in Part II of this study, the proposed 2:1 multiplexers will be integrated into SRAM as the only primary priority building block. So, cell placement of primary and second priority blocks can be accomplished forthwith after designing 2:1 multiplexer.

## IV. PART-I CONCLUSION

In this study, the Priority-Phased Decomposition-Driven (PPDD) QCA logic design methodology was proposed. In this methodology, after partitioning combinational and sequential circuits into primary-level priority blocks including 2:1 MUX and XOR gates, and secondary-level priority blocks comprising multi-input majority gates, there are three streamlined approaches which are developed to compose the desired QCA circuit using blocks from each priority level.